\documentclass[prb,twocolumn,superscriptaddress,showpacs,floatfix]{revtex4}
\usepackage{mathrsfs}
\usepackage{amsmath,amsfonts,amssymb}
\usepackage{epsfig}
\usepackage{graphicx}
\usepackage{wasysym}
\usepackage{bbm}
\usepackage{psfrag}
\usepackage{color}
\usepackage{pstool}
\usepackage{braket}
\usepackage{lmodern}
\usepackage{ulem}
\usepackage[dvips, bookmarks, colorlinks=true, plainpages = false, citecolor = blue, linkcolor = blue, urlcolor = blue, filecolor = blue]{hyperref}
\def\be{\begin{equation}}
\def\ee{\end{equation}}
\def\bea{\begin{eqnarray}}
\def\eea{\end{eqnarray}}

\begin{document}
\title{Magnetic phases of XY model with three-spin terms: 
  interplay of topology and entanglement}
\author{Rakesh Kumar Malakar}\email{rkmalakar75@gmail.com
}
\author{Asim Kumar Ghosh}
 \email{asimkumar96@yahoo.com}
\affiliation {Department of Physics, Jadavpur University, 
188 Raja Subodh Chandra Mallik Road, Kolkata 700032, India}
\begin{abstract}
  Magnetic and topological properties along with
  quantum correlations in terms of several entanglement
  measures have been investigated for an antiferromagnetic
  spin-1/2 XY model in the
  presence of transverse magnetic field and
  XZX$-$YZY type of three-spin interactions.
Symmetries of the spin Hamiltonian have been identified. 
  Under the Jordan-Wigner
  transformation, the spin Hamiltonian converted into
  spinless superconducting model with nearest neighbor hopping and
  Cooper pairing terms in addition to next nearest neighbor
  Cooper pairing potential. 
  Long range antiferromagnetic order has been studied in terms of
  staggered spin-spin correlation functions, while the topological
  orders have been characterized by winding numbers. Magnetic and topological
  phase diagrams have been prepared. 
  Faithful coexistence of magnetic and topological superconducting phases
  is found in the entire parameter regime.
  Boundaries of various quantum phases have been marked
  and positions of bicritical points have been identified. 
   \vskip 1 cm
  Corresponding author: Asim Kumar Ghosh,

  email: asimkumar96@yahoo.com
\end{abstract}
\maketitle
\section{INTRODUCTION}
Quantum phase transition (QPT) is one of the most intriguing
field of investigation in condensed matter physics for a long time,
which occurs at absolute zero temperature in interacting
many-body systems\cite{Sachdev,BKC1}. 
This type of transition is driven by the quantum fluctuations
which are governed by the strength of competitive
parameters of the system rather than
the temperature ($T$). As a result, numerous quantum phases appear
depending on the numbers of independent parameters
of the system as well as on its symmetries. 
In this context, spin models serve as potential candidates
in exhibiting various quantum phases with long range order (LRO)
those can be captured in terms of spin-spin correlation functions. 
In this series the most primitive one, the transverse-field
Ising model (TIM) has been described by a single
independent parameter and it exhibits QPT at a single point separating
phase with LRO and not\cite{Sachdev}. The next primitive member of this series 
is the transverse-field XY model (TXYM) described by two
independent parameters which exhibits
two types of QPTs along with bicritical points in the
phase diagram\cite{LSM,Katsura,Barouch1,Barouch2,Barouch3,Pfeuty}.
In line with this advancement, quantum phases
of these spin-1/2 models with additional three-spin interaction are being
studied in order to observe richer phase diagrams.
In this study, properties of quantum phases of the
TXYM in the presence of a
particular three-spin term have been investigated
in terms of magnetic LRO, topological character and
entanglement measures. 

Another recent frontier of intense research activity in condensed matter
physics is the investigation on topological properties of
quantum systems\cite{Asboth,Qi}.
Topological phase transitions are characterized by changes
in the topology of the state vectors in which nature of a definite 
phase is identified by unique value of topological invariant.
Topological insulating and superconducting phases can be understood
easily by means of the simplest one-dimensional (1D) systems, 
respectively known as SSH and Kitaev models\cite{Kitaev,SSH,Rakesh1}. 
Interestingly, TXYM becomes identical to the Kitaev
model for $p$-wave 1D superconductor upon converting spin
operator to spin-less fermions by Jordan-Wigner (JW)
transformations\cite{JW,Kitaev}. Furthermore, a faithful correspondence 
between the magnetic phase with LRO in the TXYM 
and the nontrivial topological phase in the Kitaev model
has been established. Both SSH and Kitaev models preserve the
same group of symmetries belong to the BDI class, and
their nontrivial phases have been characterized by the
topological invariant, known as winding number ($\nu$),  
with the value $\nu=1$. Nontrivial phase is associated with the
symmetry protected edge states. Charge conduction through the
edge states is immune from the back scattering, so, it is
insensitive to the local imperfection of the materials. 

Traditionally quantum correlations have been studied 
in terms of several entanglement measures\cite{Nielsen}. In order to mark the
QPTs in TIM and TXYM, measures like von Neumann entropy,  
concurrence, fidelity, quantum discord, etc have
been employed\cite{Osterloh,Osborne,Korepin1,Amico}.
Properties of nearest neighbor (NN) and next nearest
neighbor (NNN) concurrences for
TIM have been studied and their behaviors at zero and nonzero
temperatures have been obtained\cite{Osterloh}. 
Properties of concurrence of TXYM has been
studied extensively and the critical point has been identified\cite{Amico}.
The first derivative of NN concurrence and the second derivative of NNN 
concurrence exhibit logarithmic divergence at the
critical point\cite{Osterloh}. Second (first) derivative of
NN (NNN) quantum discord diverges at the critical point for
TIM\cite{Sarandy,Dillenschneider}. 
Fidelity and fidelity susceptibility exhibit sudden
changes at the critical point for TXYM\cite{Tong1}. 

In order to study more complex phase diagrams, spin models
with a variety of three-spin terms have been formulated. In this
work, TXYM with a particular form of three-spin term has been considered
and its magnetic, topological, and entanglement properties have been
investigated. Hamiltonian is introduced in Sec \ref{model}.
Symmetry of the Hamiltonian and the variety of three-spin terms have
been described. Ground state energy and low energy fermionic
excitations have been obtained in Sec \ref{JW}. Magnetic and topological
properties have been characterized in the subsections \ref{Magnetic},
and \ref{Topology}, respectively. Properties of the system in terms of
entanglement measures are presented in Sec \ref{entanglement}. 
A discussion based on these results is available 
in Sec \ref{Discussion}.
\section{Anisotropic spin-1/2 XY chain with three-spin terms}
\label{model}
Hamiltonian for the spin-1/2 XY model in the
presence of three-spin terms and transverse 
magnetic field is given by
\bea
   H&=&\sum_{j=1}^N\big[J((1+\gamma)S_j^xS_{j+1}^x
  +(1-\gamma)S_j^yS_{j+1}^y) \nonumber\\[-0.2em]
    &&\;\;+J'(S_j^xS_{j+2}^x - S_j^yS_{j+2}^y)S_{j+1}^z+hS_j^z\big].
 \label{ham}
  \eea
Here $J$ and $J'$ are the NN and three-spin exchange interaction 
strengths, and $N$ is the total number of sites.
$S_j^\alpha$, is the $\alpha\,(x,\,y,\,z)$-component of
spin-1/2 operator at the site $j$. 
$\gamma$ is the anisotropic parameter while
 $h$ is the strength of magnetic field
acting along the $z$ direction. In the absence of the
three-spin term ($J'=0$), system becomes equal to
the TXYM and in addition it reduces to several
other standard spin models in the limiting cases.
For example it is the transverse-field XX model (TXXM) at the
isotropic point, $\gamma=0$ and it becomes TIM when $\gamma=\pm 1$. 

The three-spin term introduced here is customarily referred as the
XZX$-$YZY type of interaction in regard of the positional order of
interacting spins.   
Different types of three-spin terms depending on their positions 
and symmetries for the isotropic point and
beyond can be constructed\cite{Tong1}. 
Long before Suzuki has introduced
a class of multi-spin models those are
diagonalizable by JW transformations\cite{Suzuki}. 
Subsequently, ground state phase diagram in terms of
multiple spin-liquid states of TXXM 
with XZX$+$YZY type of interaction has been studied\cite{Japaridze}. 
Properties of thermodynamic quantities of XXZ model 
with three different types of three-spin terms,
XZY$-$YZX, YXZ$-$ZXY, and ZYX$-$XYZ, have been investigated\cite{Lou}.
Revival structure in a Loschmidt echo is observed in
the XX model with XZX$+$YZY term and in the presence of
staggered magnetic field\cite{Jafari}. 
It has been shown that TXXM model with XZX$+$YZY term
can be mapped on to the XZY$-$YZX upon spin rotation
\cite{Derzhko1,Derzhko2}. 

For the TXYM with XZX$+$YZY term quantum
criticality has been studied in terms of
fidelity susceptibility\cite{Cheng1}. Behavior of thermal
quantum and classical correlations for TXYM with XZX$+$YZY
term has been investigated\cite{Lin}. Recently, magnetic
and topological phase diagrams of the same model have been
investigated extensively\cite{Rakesh2}. Coexistence of
magnetic LRO with the classical correlation and quantum discord
has been observed when $T=0$. Unlike the TXXM, TXYM with
XZX$+$YZY term is different from that with
XZY$-$YZX term\cite{Tong1}. Dynamical entanglement of two
spin-qubits coupled to TXYM with XZY$-$YZX term
has been studied\cite{Cheng2}. QPTs and existence of
chiral phase has been demonstrated for the same model\cite{Tong1}. 
In this study, TXYM with the XZX$-$YZY type of three-spin interactions
has been considered and investigated extensively. 

The corresponding Hamiltonian (Eq. \ref{ham}) lacks the SU(2) symmetry
when $\gamma \ne 0$ and $J' \ne 0$, 
since $[H,\,S_{\rm T}^\alpha]\ne 0$, where
$S_{\rm T}^\alpha$ is the $\alpha$-component of
the total spin, $\boldsymbol S_{\rm T}=\sum_{j=1}^N \boldsymbol S_j$.
Similarly, it does not preserve  
the time-reversal symmetry (TRS) as long as
$J' \ne 0$ and $h \ne 0$. 
Nevertheless, Hamiltonian 
possesses a number of symmetries as described below. 
\begin{subequations}
\begin{align}
&R_z(\pi) H(J,J',\gamma,h) R_z^\dag(\pi)=H(J,J',\gamma,h),\\[0.4em]
&R_z(\pi/2)H(J,J',\gamma,h) R^\dag_z  (\pi/2)=H(J,-J',-\gamma,h),\\[0.4em]
&R_{x,y} (\pi)H(J,J',\gamma,h)R_{x,y}^\dag(\pi)=H(J,-J',\gamma,-h),\\[0.4em]
&R'_z (\pi)H(J,J',\gamma,h) R'^\dag_z  (\pi)=H(-J,J',\gamma,h),\\[0.4em]
&RH(J,J',\gamma,h) R^\dag=-H(J,J',\gamma,h),\\[0.4em]
&R'H(J,J',\gamma,h) R'^\dag=H(J,J',-\gamma,-h), 
\end{align}
\label{symmetry}
\end{subequations}
where $R_\alpha(\theta)=\prod_{j=1}^Ne^{i\theta S_j^\alpha}$ is
the rotation operator about the $\alpha$-axis by an angle
$\theta$, and $R'_\alpha(\theta)=\prod_{j=1,3,5}^Ne^{i\theta S_j^\alpha}$.
$R$ and $R'$ are the products of two different rotations, where 
$R=R'_z(\pi)R_{x,y} (\pi)$, and $R'=R_z(\pi/2)R_{x,y} (\pi)$. 
The first relation (Eq (\ref{symmetry}a)) corresponds to the $Z_2$
symmetry, while 
the second to last one (Eq (\ref{symmetry}e)) act like  the
particle-hole symmetry of the system. As a consequence,
energy spectrum acquires the inversion symmetry about the zero.
It also implies the existence zero energy states in pairs.
The relation in Eq (\ref{symmetry}d) holds for even value of $N$.
The antiferromagnetic (AFM) Hamiltonian will be
converted to ferromagnetic (FM) one under this transformation. 
Obviously these relations will be different for
different types of three-spin terms. For example,
in case of TXYM with XZX$+$YZY term, all the
relations hold except the second one (Eq (\ref{symmetry}b)). The second relation
for this case is
$R_z (\pi/2)H(J,J',\gamma,h) R^\dag_z  (\pi/2)=H(J,J',-\gamma,h)$. 
This difference ultimately leads to a huge deviations in the
magnetic and topological properties at the end in such an 
amplitude that no connections between them could be established.

In this study, magnetic and topological properties of this
model have been investigated extensively.
Long range correlation of AFM order
in a system can be studied by evaluating the value of 
N\'eel (staggered spin-spin) correlation functions
 ${\mathcal C}_{\textrm{N\'eel}}^\alpha (n)$.  
The expressions of the staggered magnetizations and correlations are given by  
\bea
m_{\textrm{N\'eel}}^\alpha
&=&\frac{1}{N}\sum_{j=1}^N (-1)^j \langle S_j^\alpha\rangle,\\
{\mathcal C}_{\textrm{N\'eel}}^\alpha (n)&=&\frac{1}{N}\sum_{j=1}^N(-1)^n\langle  S_j^\alpha \,S_{j+n}^\alpha\rangle,
\label{CF}
\eea
where the number $n$ denotes the separation between two spins
between which the correlation is to be measured.
Expectation value $\langle \cdot \rangle$ is
 evaluated in the ground state.
Correlation functions hold the relations, 
${\mathcal C}_{{\textrm {N\'eel}}}^x(n)\ne{\mathcal C}_{{\textrm {N\'eel}}}^y (n)
\ne{\mathcal C}_{\textrm {N\'eel}}^z (n)$ as long as $\gamma \ne 0$, 
whereas ${\mathcal C}_{{\textrm {N\'eel}}}^x(n)={\mathcal C}_{{\textrm {N\'eel}}}^y (n)$,
when $\gamma=0$, owing to the symmetries of the system.
For the  $Z_2$ symmetry of the Hamiltonian, staggered magnetization,
$m_{\textrm{N\'eel}}^\beta,\;\beta=x,\,y$, must vanish albeit the
correlations ${\mathcal C}_{\textrm{N\'eel}}^\beta (n)$ and
the $z$-component of the magnetization may survive as long as the
ground state preserve that symmetry.
It happens as a consequence of the relations:
 $R_z(\pi) \,S_j^\beta S_{j+n}^\beta \,R_z^\dag(\pi)=S_j^\beta S_{j+n}^\beta,$ 
and $R_z(\pi) \,S_j^\beta \,R_z^\dag(\pi)=-S_j^\beta,$ but
$R_z(\pi) \,S_j^z \,R_z^\dag(\pi)=S_j^z$. It is true for both 
TIM and TXYM, but not for the Ising model where ground state breaks the
$Z_2$ symmetry, as a result, both $m_{\textrm{N\'eel}}^\beta$, and 
${\mathcal C}_{\textrm{N\'eel}}^\beta (n)$ survive simultaneously,
leading to the LRO.

It is worth mentioning at this point that
all the symmetries described in
the Eqs. (2a)-(2f) evaluated for the Hamiltonian (Eq. 1), 
also hold for the Ising, XY and TXYM, since all these
models can be derived form the Hamiltonian.
The Ising model, $H(\gamma=\pm 1,J'=0,h=0)$, exhibits LRO,
but no phase transition occurs since there is no term
which could destroy the LRO. However the TIM,
$H(\gamma=\pm 1,J'=0,h\ne 0)$, exhibits
LRO when $-1<h/J<1$, but no LRO 
for either $h/J>1$, or $h/J>-1$, leading to the phase transition at the points 
$h/J=\pm 1$. On the other hand,
the TXYM, $H(\gamma \ne 0,J'=0,h\ne 0)$, exhibits
two different long-range ordered phases for $-1<h/J<1$, when $\gamma>0$, and
$\gamma<0$. However, both the ordered phases vanish as soon as the
the value of magnetic filed is  $h/J>1$, or $h/J>-1$, leading to the
disordered phase.
It means the points $\gamma=0$ and $h/J=\pm 1$, is the meeting points of
three different phases. 
In order to study magnetic and topological properties,
the Hamiltonian (Eq. \ref{ham}) has been expressed
in terms of spin-less fermions as discussed in the
following section.
\section{Hamiltonian in Jordan-Wigner fermion}
\label{JW}
Dispersion relations, ground state energy and
spin-spin correlation functions of $H$ have been obtained
analytically in terms of fermions after converting the
spin components using the JW 
transformations\cite{JW}:  
$S_j^+=c_j^\dag\prod_{l=1}^{j-1}(1-2\,n_l),\;
S_j^-=\prod_{l=1}^{j-1}(1-2\,n_l)\,c_j,\;
S_j^z=n_j-\frac{1}{2}$,
where $n_l=c_l^\dag c_l$, is the number operator
for the JW fermions. The transformed Hamiltonian 
lookes like  
\bea
   H&=&\!\frac{1}{2}\sum_j\!\bigg[J\left(c_j^\dag c_{j+1}\!+\!c_{j+1}^\dag c_j\right)
     -\frac{J'}{2}\left(c_j^\dag c_{j+2}^\dag+c_{j+2}c_j\right)\nonumber\\[-0.2em]
     && \;\;  +J\gamma\left(c_{j+1}^\dag c_j^\dag+ c_j c_{j+1}\right)
     +2h\left(c_j^\dag c_{j}-\frac{1}{2}\right)\bigg],
 \label{hamJW}
\eea
where it comprises of single particle terms
and similar to the $p$-wave superconducting
Hamiltonian which allows both NN and NNN Cooper pairings,
and NN hopping\cite{Kitaev}. 
Here $J$ plays the role of NN hopping
parameter while and $J'/2$ and $\gamma J$ are the
respective NN and NNN superconducting potentials where 
$h$ is the chemical potential. 
As a result,  Hamiltonian assumes the
Bogoliubov-de Gennes (BdG) form 
under the Fourier transformation, 
$c_j=\frac{1}{\sqrt N}\sum_{k}c_{k}\,e^{-i{k}j},$
where
\be
H=\sum_{k}\left[\psi_{k}^\dag\, \mathcal H_{k}\,\psi_{k}+\epsilon_{k}-\frac{h}{2}\right],
\ee
 $\psi_{k}^\dag=\left[c^\dag_{k}\;\,c_{-k}\right]$,  
$\epsilon_{k}=(J\cos{(k)}+h)/2$, 
$\mathcal H_{k}=\boldsymbol g \cdot \boldsymbol \sigma$,
with Pauli matrices, 
 $\boldsymbol \sigma=(\sigma_x,\sigma_y,\sigma_z)$, 
$\boldsymbol g=(0,g_y,g_z)$, and 
$g_y=\Delta_{k},$ $g_z=\epsilon_{k}$, 
 $\Delta_{k}=\frac{1}{2}J\gamma\sin{(k)}-\frac{1}{4}J'\sin{(2k)}$.
Dispersion relation and the ground state energy per site
are respectively given by 
\be E_{k}=\sqrt{\epsilon_{k}^2+|\Delta_{k}|^2},\ee 
and 
\be
E_{\rm G}=-\frac{1}{2\pi}\int_{-\pi}^{+\pi}E_{k} d{k}.
\label{Eq-Eg}
\ee
This superconducting system favors the formation of Cooper pair
between parallel spins and so 
there is a gap above the ground state energy as long as
the superconducting parameter $\Delta_{k} \ne 0$.
However, the energy gap vanishes over the boundary line
separating different phases which will be discussed in the
subsequent sections. So, these boundary lines in the parameter space
can be identified easily by the equation, $E_{k}=0$.
Which ultimately leads to equation, $\gamma=-hJ'/J^2$, with the
restriction that when the minimum of $E_{k}$ does not occur on
the points $k=0,\, \pi$.
Further energy gap vanishes over two horizontal lines,
$h/J=\pm 1$ in the $h$-$J'$ plane, when $k=\pi$ and
$k=0$, respectively. Ultimately different quantum phases will be
separated by the rectangular hyperbolas, $(h/J)(J'/J)=-\gamma$,
in the $h$-$J'$ plane bounded by the horizontal lines,
$h/J=\pm 1$. The meeting points of the hyperbolas and the
horizontal lines lead to two different bicritical points, where
three different phases coexist.
The idea of multicriticality is introduced before in order to
study the critical behavior at the transition point in case of 
thermodynamic phase transition, where dimension of
order parameter plays the crucial role. The existence
of bicritical points is experimentally found in crystalline $^4$He,
mixed magnetic crystals, {\it e. g.}, (MN, Fe)WO$_4$ or
Fe(Pd, Pt)$_3$, perovskite crystals, etc\cite{Elliott}. 
In order to identify the character of quantum phases
in different regions in the $h$-$J'$ plane, magnitude of the
correlation functions and the value of the winding number have been
determined in the following two subsections. 
\subsection{Magnetic phases}
\label{Magnetic}
In order to find the magnitude of correlation functions, they are
expressed in terms of spinless fermionic operators \cite{LSM,Barouch1,Barouch2}, 
\bea {\mathcal C}_{\textrm {N\'eel}}^x(l\!-\!m)&=&\frac{1}{4}\langle B_lA_{l+1}B_{l+1}
\cdots A_{m-1}B_{m-1}A_m\rangle,\nonumber\\[0.3em]
{\mathcal C}_{\textrm {N\'eel}}^y(l\!-\!m)&=&\frac{(-1)^{m-l}}{4}\langle A_lB_{l+1}A_{l+1}
\cdots B_{m-1}A_{m-1}B_m\rangle,\nonumber\\[0.3em]
{\mathcal C}_{\textrm {N\'eel}}^z(l\!-\!m)&=&\frac{1}{4}\langle A_lB_{l}A_{m}B_{m}\rangle,
\nonumber\eea
where $A_j=c_j^\dag+c_j$, and $B_j=c_j^\dag-c_j$.
Employing the Wick's theorem and using the
identity $\langle A_lA_m\rangle=\langle B_lB_m\rangle=\delta_{lm}$,
as well as the translational invariance, 
they are expressed in terms of determinant of Toeplitz matrices, 
\be{\mathcal C}_{\textrm {N\'eel}}^x(n)=\frac{1}{4}\left|
\begin{array}{cccc}G_{-1}&G_{-2}&\cdots&G_{-n}\\[0.4em]
  G_{0}&G_{-1}&\cdots&G_{-n+1}\\[0.2em]
  \vdots&\vdots&\ddots&\vdots\\[0.2em]
   G_{n-2}&G_{n-3}&\cdots&G_{-1}\end{array}
\right|,\label{JWCx}\ee
\be{\mathcal C}_{\textrm {N\'eel}}^y(n)=\frac{1}{4}\left|
\begin{array}{cccc}G_{1}&G_{2}&\cdots&G_{n}\\[0.4em]
  G_{0}&G_{1}&\cdots&G_{n-1}\\[0.2em]
  \vdots&\vdots&\ddots&\vdots\\[0.2em]
   G_{-n+2}&G_{-n+3}&\cdots&G_{1}\end{array}
\right|,
\label{JWCy}
\ee
\be{\mathcal C}_{\textrm {N\'eel}}^z(n)=m_z^2-G_{n}G_{-n}/4,\label{JWCz}\ee
where matrix elements are given
in terms of the ground state expectation values,
$G_{n}=\langle B_{n+l}A_{l} \rangle$ and the $z$-component of
magnetization $m_z=\langle S^z\rangle$.
Now imposing the PBC and for $T=0$, 
\bea m_z&=&\frac{1}{2\pi}\int_0^\pi\! d{k}\,\frac{\epsilon_{k}}{E_{k}},
\nonumber \\[0.5em]
G_{n}&=&\frac{1}{\pi}\int_0^\pi\!\! d{k}\,
\frac{\epsilon_{k}\cos{({k}n)}-\Delta_{k}\sin{({k}n)}}
     {E_{k}}.\nonumber \eea
     
\begin{figure}[h]
\psfrag{a}{(a)}
\psfrag{b}{(b)}
\psfrag{M}{\color{blue}$m_z$}
\psfrag{dM}{\hskip 0.1 cm \color{red}$dm_z/dh$}
\psfrag{p1}{$J=1,\,J'=3,\,\gamma=2.0$}
\psfrag{p2}{$J=1,\,J'=3,\,\gamma=-2.0$}
\psfrag{h}{$h/J$}
\psfrag{0}{0}
\psfrag{1.0}{$1.0$}
\psfrag{-1.0}{$-1.0$}
\psfrag{2}{2}
\psfrag{-1.0}{$-1.0$}
\psfrag{-3}{$-3$}
\psfrag{-2}{$-2$}
\psfrag{-4}{$-4$}
\psfrag{3}{3}
\psfrag{4}{4}
\psfrag{-0.5}{$-0.5$}
\psfrag{-1.5}{$-1.5$}
\psfrag{1.5}{$1.5$}
\psfrag{0.0}{$0.0$}
\psfrag{0.5}{$0.5$}
\psfrag{-0.8}{$-0.9$}
\psfrag{-0.9}{$-0.9$}
\psfrag{0.15}{$0.15$}
\psfrag{0.20}{$0.20$}
\psfrag{0.25}{$0.25$}
\psfrag{d}{Disordered}
\psfrag{Cx}{${\mathcal C}_{\textrm{N\'eel}}^x\ne 0$}
\psfrag{Cy}{${\mathcal C}_{\textrm{N\'eel}}^y\ne 0$}
\includegraphics[width=250pt]{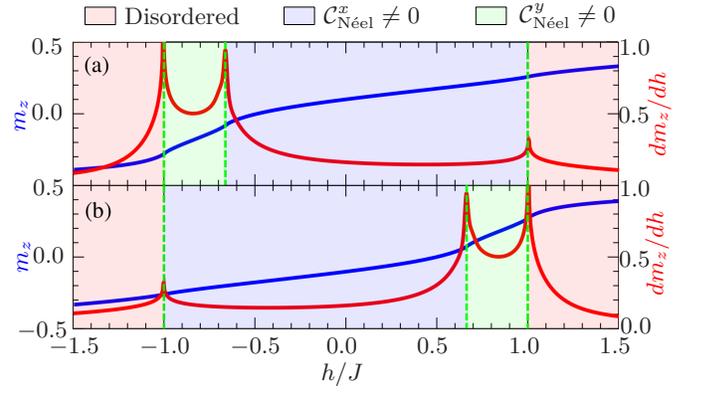}
\caption{Variation of magnetization (blue) and its first order derivative
  with respect to $h$ (red) for $J'=3,$ when $\gamma=2.0$ (a)
  and $\gamma=-2.0$ (b). Vertical dashed lines (green) indicate the
positions of QPTs.}
 \label{magnetization}
\end{figure}

\begin{figure}[h]
\psfrag{d}{\color{white} Disordered}
\psfrag{a}{\color{white} ${\mathcal C}_{\textrm{N\'eel}}^x\ne 0$}
\psfrag{b}{\color{white}${\mathcal C}_{\textrm{N\'eel}}^y\ne 0$}
\psfrag{dM}{\hskip 0.1 cm \color{red}$dm_z/dh$}
\psfrag{h}{$h/J$}
\psfrag{jp}{$J'/J$}
\psfrag{0}{0}
\psfrag{1.0}{$1.0$}
\psfrag{-1.0}{$-1.0$}
\psfrag{2}{$2$}
\psfrag{1}{$1$}
\psfrag{-1}{$-1$}
\psfrag{-1.0}{$-1.0$}
\psfrag{-3}{$-3$}
\psfrag{-2}{$-2$}
\psfrag{-4}{$-4$}
\psfrag{3}{$3$}
\psfrag{4}{$4$}
\psfrag{5}{$5$}
\psfrag{-5}{$-5$}
\psfrag{-0.5}{$-0.5$}
\psfrag{-1.5}{$-1.5$}
\psfrag{1.5}{\hskip 0.0 cm $1.5$}
\psfrag{0.0}{$0.0$}
\psfrag{0.5}{$0.5$}
\psfrag{0.6}{$0.6$}
\psfrag{0.8}{$0.8$}
\psfrag{-0.9}{$-0.9$}
\psfrag{1.6}{$1.6$}
\psfrag{1.4}{$1.4$}
\psfrag{1.2}{$1.2$}
\psfrag{0.2}{$0.2$}
\psfrag{0.4}{$0.4$}
\includegraphics[width=250pt,angle=0]{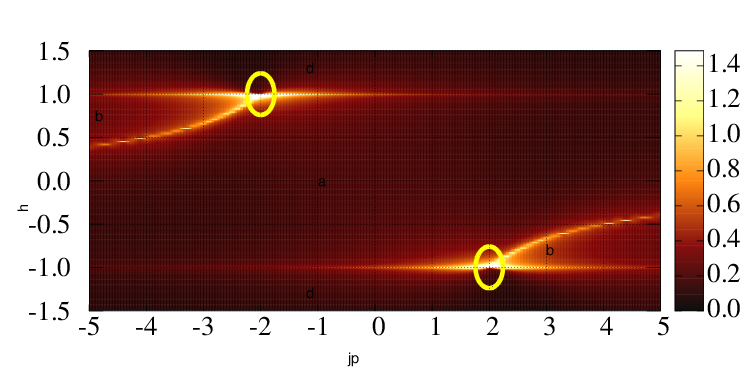}
\caption{Density plot for the first order derivative of magnetization 
  with respect to magnetic field, $dm_z/dh$, 
  in the $h$-$J'$ plane is shown for $\gamma=2$. 
  Upper and lower bicritical
  points are enclosed within yellow circles. }
 \label{derivative-magnetization}
\end{figure}

\begin{figure}[h]
\psfrag{d}{\color{white} Disordered}
\psfrag{a}{\color{white} ${\mathcal C}_{\textrm{N\'eel}}^x\ne 0$}
\psfrag{b}{\color{white}${\mathcal C}_{\textrm{N\'eel}}^y\ne 0$}
\psfrag{h}{$h/J$}
\psfrag{gm}{$\gamma$}
\psfrag{0}{0}
\psfrag{1.0}{$1.0$}
\psfrag{-1.0}{$-1.0$}
\psfrag{1}{$1$}
\psfrag{2}{$2$}
\psfrag{-1.0}{$-1.0$}
\psfrag{-3}{$-3$}
\psfrag{-2}{$-2$}
\psfrag{-1}{$-1$}
\psfrag{-4}{$-4$}
\psfrag{3}{$3$}
\psfrag{4}{4}
\psfrag{-0.5}{$-0.5$}
\psfrag{-1.5}{$-1.5$}
\psfrag{1.5}{\hskip 0.0 cm $1.5$}
\psfrag{0.0}{$0.0$}
\psfrag{0.5}{$0.5$}
\psfrag{0.6}{$0.6$}
\psfrag{0.8}{$0.8$}
\psfrag{-0.9}{$-0.9$}
\psfrag{1.6}{$1.6$}
\psfrag{1.4}{$1.4$}
\psfrag{1.2}{$1.2$}
\psfrag{0.2}{$0.2$}
\psfrag{0.4}{$0.4$}
\includegraphics[width=250pt,angle=0]{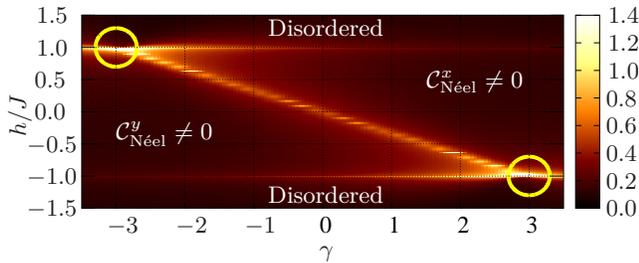}
\caption{Density plot for the first order derivative of magnetization 
  with respect to magnetic field, $dm_z/dh$, 
  in the $h$-$\gamma$ plane is shown for $J'=3$. 
  Upper and lower bicritical
  points are enclosed within yellow circles. }
 \label{derivative-magnetization-2}
\end{figure}

Variation of $m_z$ and its derivative with respect to the
magnetic field, $dm_z/dh$, are shown in blue and red lines,
respectively, in Fig. \ref{magnetization},  
for $J'=3,$ when $\gamma=2$ (a)
and $\gamma=-2$ (b).
In every figure, the NN exchange interaction strength
will be assumed as the unit of energy, i.e. $J = 1$. 
Vertical dashed lines (green) indicate the
positions of QPTs, where $dm_z/dh$ exhibits sharp peaks.
Apart from the points $h/J=\pm 1$, other points of QPTs can be marked by
$h/J=\mp 2/3$, for $\gamma=\pm 2$.
The positions of the QPTs about $h=0$ can be understood in terms of the
symmetry of the magnetization, $m_z(J,\pm \gamma,J',\pm h)
=m_z(J,\mp \gamma, J',\mp h)$  according to the relation
Eq (\ref{symmetry}f). 
Density plot of the $dm_z/dh$  in the $h$-$J'$ plane is
shown for $\gamma=2$, in Fig. \ref{derivative-magnetization}.
Bright lines clearly indicate the position of QPTs.
Besides the horizontal lines $h/J=\pm 1$, curves are represented by
$hJ'=2$. Position of the upper and lower bicritical
points are given by $(1,\,-2)$ and $(-1,\,2)$, respectively, 
those are shown within the yellow circles.
The symmetry of the QPTs in the $h$-$J'$ plane corresponds to that of the
magnetization, $m_z(J,\gamma,\pm J',\pm h)
=m_z(J,\gamma,\mp J',\mp h)$, according to the relation
Eq (\ref{symmetry}b). These results are in accordance to
the plot as shown in Fig. \ref{magnetization} (a). 
Similarly density plot of the $dm_z/dh$  in the $h$-$\gamma$ plane is
shown for $J'=3$, in Fig. \ref{derivative-magnetization-2}. 
QPT lines and location of bicritical points are shown. 


\begin{figure*}[t]
\psfrag{a}{\color{white}(a)}
\psfrag{b}{\color{white}(b)}
\psfrag{c}{ \color{white}(c)}
\psfrag{d}{\color{white}(d)}
\psfrag{e}{ \color{white}(e)}
\psfrag{f}{ \color{white}(f)}
\psfrag{g}{ \color{white}(g)}
\psfrag{h1}{ \color{white}(h)}
\psfrag{i}{ \color{white}(i)}
\psfrag{j}{ \color{white}(j)}
\psfrag{g-2}{\color{white}$\gamma=-2$}
\psfrag{g-1}{\color{white}$\gamma=-1$}
\psfrag{g0}{\color{white}$\gamma=0$}  
\psfrag{g1}{\color{white}$\gamma=1$}
\psfrag{g2}{\color{white}$\gamma=2$}  
\psfrag{k}{\color{white} (k)}
\psfrag{l}{\color{white} (l)}
\psfrag{m}{\color{white} (m)}
\psfrag{n}{\color{white} (n)}
\psfrag{o}{\color{white} (o)}
\psfrag{h}{$h/J$}
\psfrag{jp}{$J'/J$}
\hskip -1.5 cm
\begin{minipage}{0.33\textwidth}
  \includegraphics[width=210pt,angle=0]{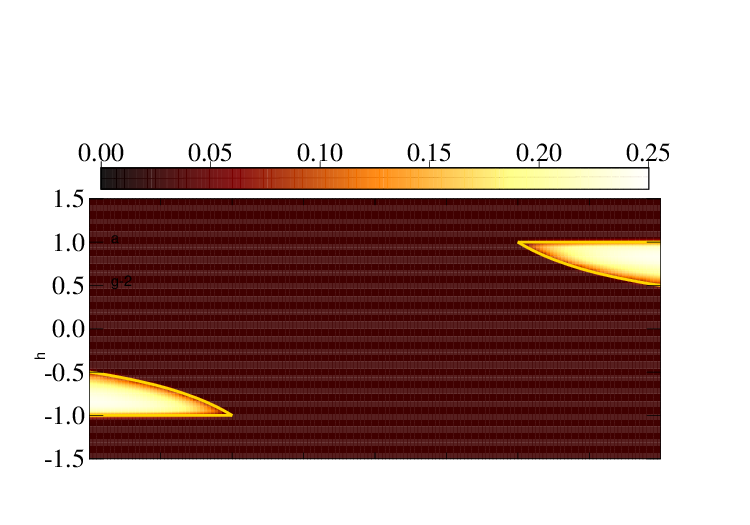}
    \end{minipage}\hskip 0.1cm
  \begin{minipage}{0.33\textwidth}
  \includegraphics[width=210pt,angle=0]{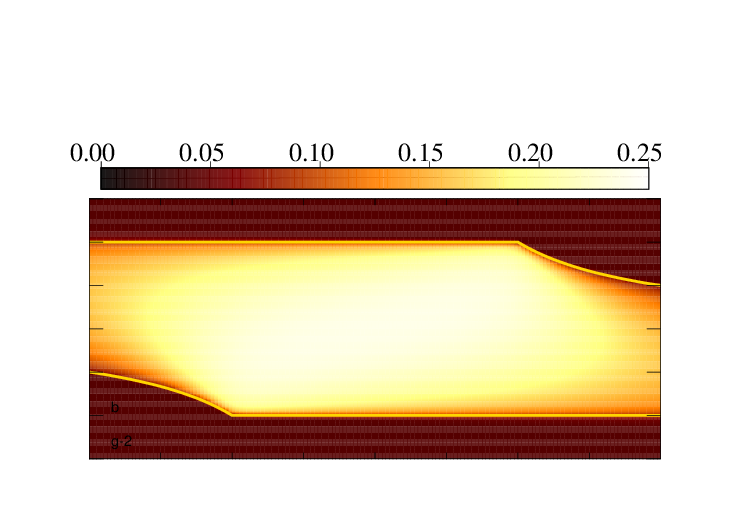}
  \end{minipage}\hskip 0.1cm
  \begin{minipage}{0.33\textwidth}
\includegraphics[width=210pt,angle=0]{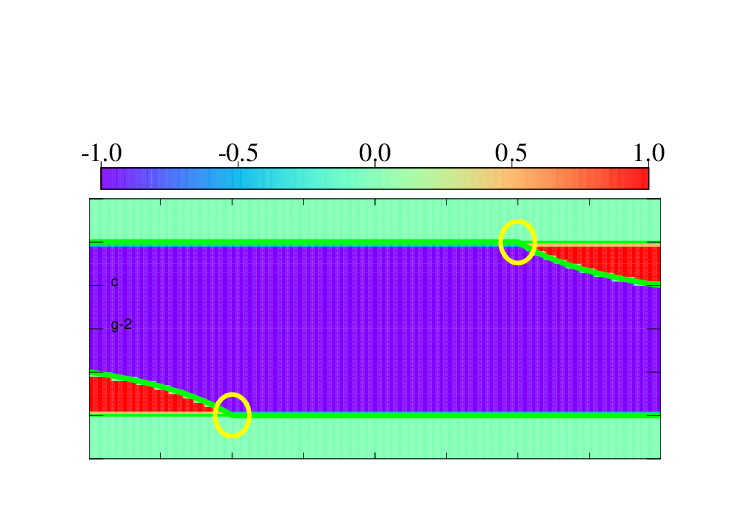}      
  \end{minipage}
  \vskip -0.97 cm
  \hskip -1.5 cm
\begin{minipage}{0.33\textwidth}
  \includegraphics[width=210pt,angle=0]{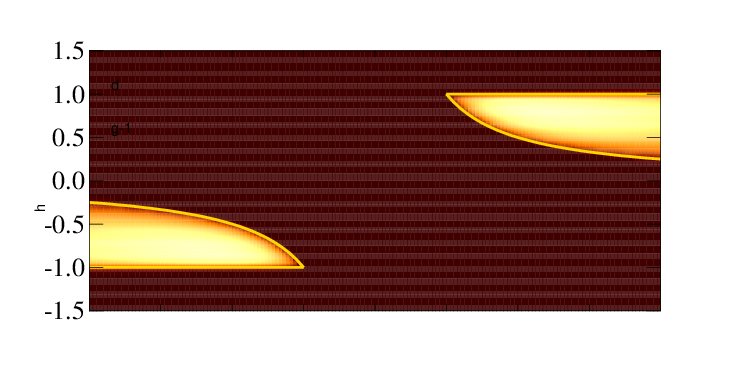}
    \end{minipage}\hskip 0.1cm
  \begin{minipage}{0.33\textwidth}
  \includegraphics[width=210pt,angle=0]{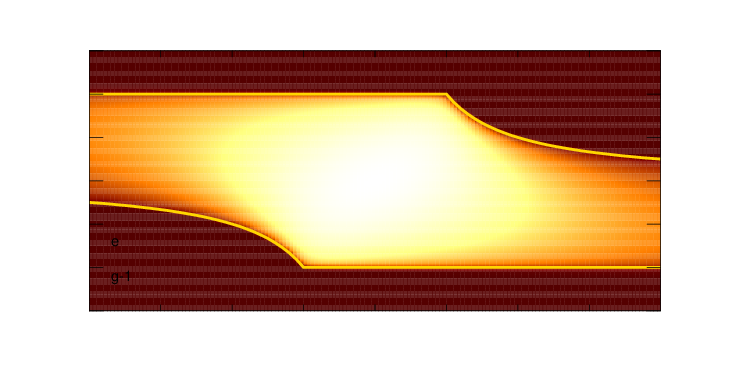}
  \end{minipage}\hskip 0.1cm
  \begin{minipage}{0.33\textwidth}
\includegraphics[width=210pt,angle=0]{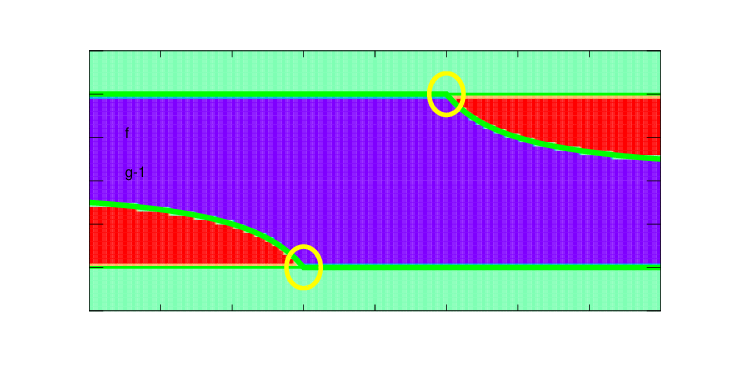}    
  \end{minipage}
    \vskip -0.97 cm
  \hskip -1.5 cm
\begin{minipage}{0.33\textwidth}
  \includegraphics[width=210pt,angle=0]{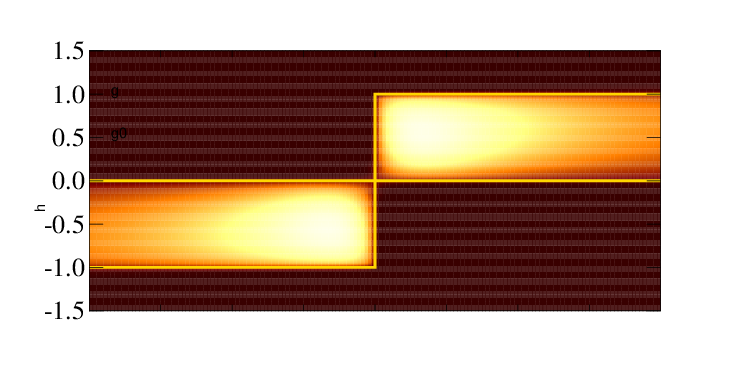}
    \end{minipage}\hskip 0.1cm
  \begin{minipage}{0.33\textwidth}
  \includegraphics[width=210pt,angle=0]{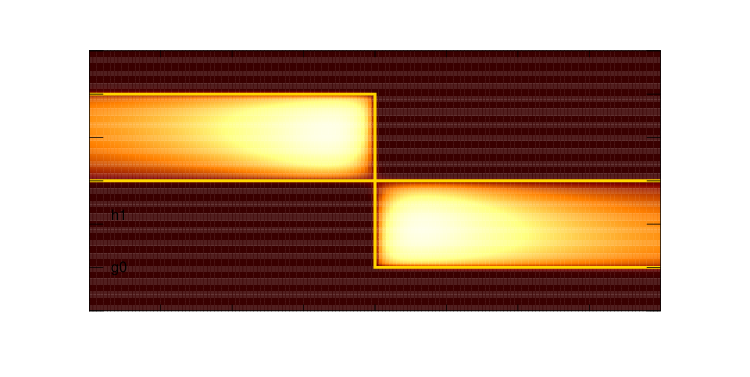}
  \end{minipage}\hskip 0.1cm
  \begin{minipage}{0.33\textwidth}
\includegraphics[width=210pt,angle=0]{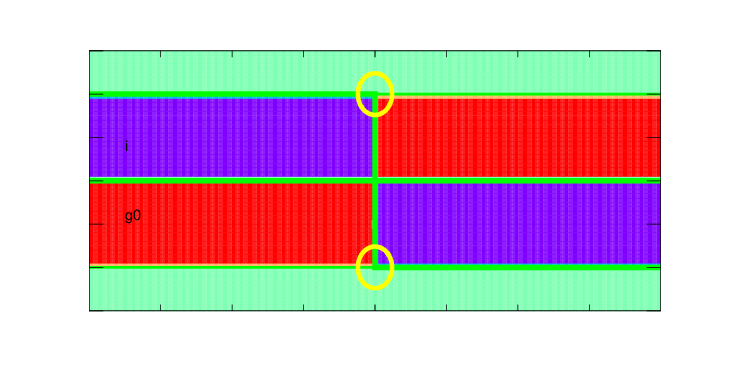}     
  \end{minipage}
  \vskip -0.97 cm
\hskip -1.5 cm
\begin{minipage}{0.33\textwidth}
  \includegraphics[width=210pt,angle=0]{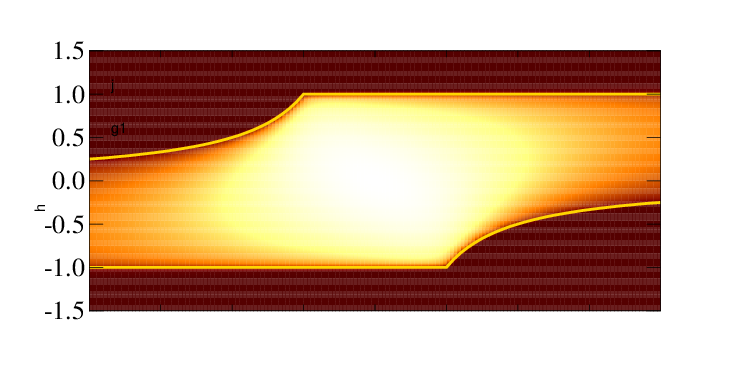}
    \end{minipage}\hskip 0.1cm
  \begin{minipage}{0.33\textwidth}
  \includegraphics[width=210pt,angle=0]{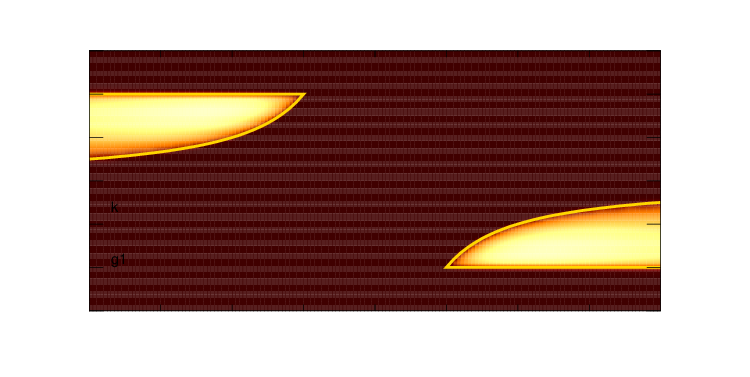}
  \end{minipage}\hskip 0.1cm
  \begin{minipage}{0.33\textwidth}
\includegraphics[width=210pt,angle=0]{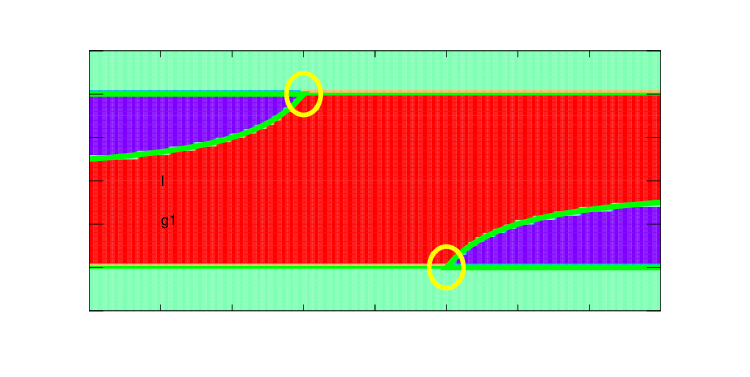}    
  \end{minipage}
  \vskip -0.97 cm
\hskip -1.5 cm
\begin{minipage}{0.33\textwidth}
  \includegraphics[width=210pt,angle=0]{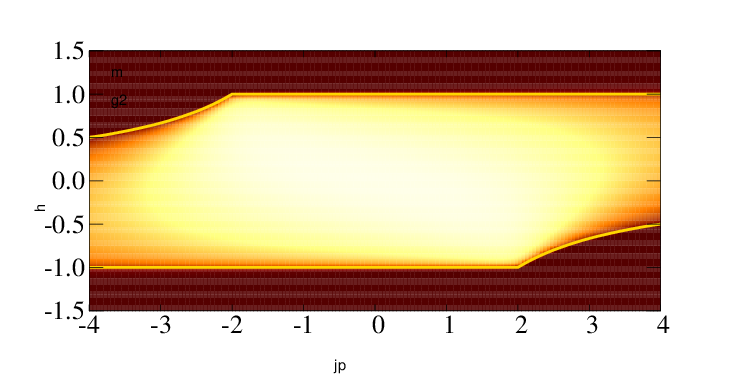}
    \end{minipage}\hskip 0.1cm
  \begin{minipage}{0.33\textwidth}
  \includegraphics[width=210pt,angle=0]{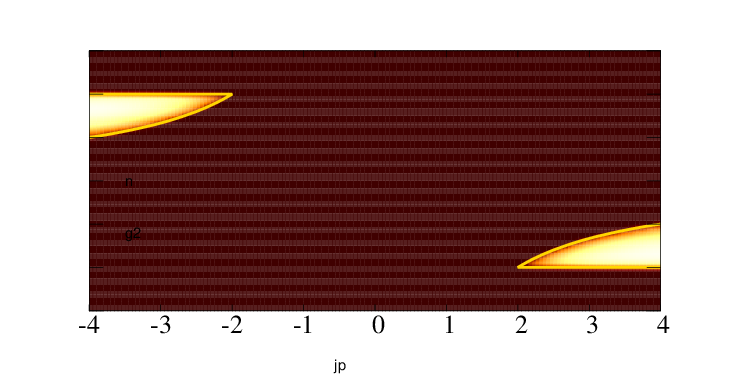}
  \end{minipage}\hskip 0.1cm
  \begin{minipage}{0.33\textwidth}
\includegraphics[width=210pt,angle=0]{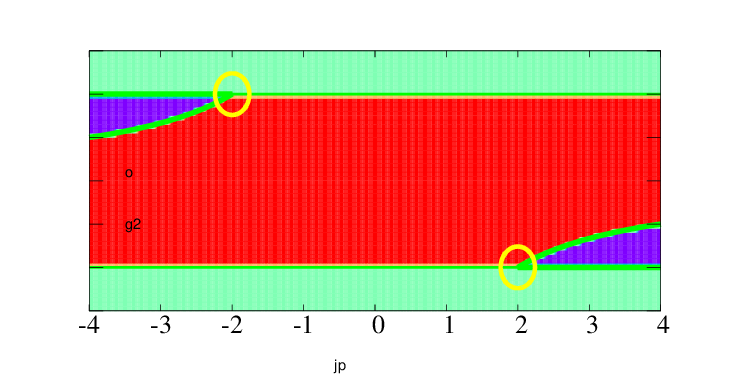}
  \end{minipage}
  \caption{Density plot for the variation of correlation functions
    and winding numbers in the $h$-$J'$ plane, for 
   $\gamma=\pm 2,\pm 1,0$. Correlations ${\mathcal C}_{\textrm {N\'eel}}^x(n)$,
    ${\mathcal C}_{\textrm {N\'eel}}^y(n)$ with $n=100$
    and winding number have been shown in the
    first ((a), (d), (g), (j), (m)), second ((b), (e), (h), (k), (n)),
    and third ((c), (f), (i), (l), (o)) vertical panel, respectively.
    Each row is drawn for a definite
  value of $\gamma$. Solid lines indicate the boundary
  of different phases. Upper and lower bicritical
  points are enclosed within yellow circles in the third vertical panel.}
\label{magnetic-topology}
\end{figure*}

Correlation functions ${\mathcal C}_{\textrm {N\'eel}}^x(n)$, and 
${\mathcal C}_{\textrm {N\'eel}}^y(n)$ for $n=100$, have been shown in the
Fig. \ref{magnetic-topology} for five different values of
$\gamma =0,\pm 1, \pm 2$. The density plots (a), (d), (g), (j) and (m)
are for the ${\mathcal C}_{\textrm {N\'eel}}^x(100)$,
while (b), (e), (h), (k) and (n)
are for the ${\mathcal C}_{\textrm {N\'eel}}^y(100)$.
Coordinates of the bicritical points in the $h$-$J'$ plane is given by
$(\pm 1,\,\mp \gamma)$. 
In this model, it is the meeting point for three
different phases among them two have different
LROs and one is disordered. The LROs are characterized by the
nonzero values of ${\mathcal C}_{\textrm {N\'eel}}^x(n)$
and ${\mathcal C}_{\textrm {N\'eel}}^y(n)$, as shown in
first and second vertical panels of
Fig. \ref{magnetic-topology}. The transitions across
the straight lines $h/J=\pm 1$, in $h$-$J'$ plane
is termed as the Ising transition while
those across the curve $(h/J)(J'/J)=-\gamma$, is known as
anisotropy transition, and they meet
at the bicritical points\cite{BKC1}.

Configuration of the phase diagrams in $h$-$J'$ plane can be understood
by means of the symmetry relations those are derived
before in Eqs (\ref{symmetry}a)-(\ref{symmetry}f). 
For arbitrary $\gamma$, the relation ${\mathcal C}_{\textrm {N\'eel}}^\beta(n,J,\gamma,\pm J',\pm h)
={\mathcal C}_{\textrm {N\'eel}}^\beta(n,J,\gamma,\mp J',\mp h)$,
$\beta=x,\,y$, holds for any values of $n$,
which corresponds to the symmetry in Eq (\ref{symmetry}c). 
Similarly, ${\mathcal C}_{\textrm {N\'eel}}^x(n,J,\pm \gamma,\pm J',h)
={\mathcal C}_{\textrm {N\'eel}}^y(n,J,\mp \gamma,\mp J',h)$,
in accordance to the symmetry, Eq (\ref{symmetry}b). 
In addition,  ${\mathcal C}_{\textrm {N\'eel}}^x(n,J,\pm \gamma,J',\pm h)
={\mathcal C}_{\textrm {N\'eel}}^y(n,J,\mp \gamma, J',\mp h)$
corresponds to the relation  Eq (\ref{symmetry}f).
Following the symmetry transformation in Eq (\ref{symmetry}d),
FM spin-spin correlation,
${\mathcal C}_{\textrm FM}^\alpha$ can be
derived from the staggered one, ${\mathcal C}_{\textrm {N\'eel}}^\alpha$.
At the same time, the uniform magnetization,
$m_\beta,\;\beta=x,\,y$, vanishes since the Hamiltonian,
Eq (\ref{ham}), preserves the $Z_2$ symmetry (Eq (\ref{symmetry}a)). 
For the limiting cases, either $\gamma\gg 1$,
or $J'\gg 1$, ${\mathcal C}_{\textrm {N\'eel}}^\beta(n)=0$,
since $G_{n}=0$ for any values of $n$. Which means
N\'eel correlation breaks down when the NN or NNN 
parallel spin Cooper-pairing is very strong.

These results are drastically different from those
obtained in case of TXYM with XZX+YZY. 
The main difference is that for TXYM with XZX-YZY, LRO diminishes with the
increase of magnetic field and vanishes 
beyond $h/J>1$, and $h/J<-1$, which is
identical to the TIM and TXYM. However, for
TXYM with XZX+YZY, LRO survives at any value of the 
magnetic field\cite{Rakesh2}.
In this model, only ${\mathcal C}_{\textrm {N\'eel}}^x$ is non-zero
only for $\gamma>0$, while  ${\mathcal C}_{\textrm {N\'eel}}^y$ is non-zero
only for $\gamma<0$. The same picture is found in TXYM.
As a result, the bicritical points always exist for $\gamma=0$, 
in TXYM both in the presence and absence of XZX+YZY type of
interaction\cite{Rakesh2}.
In contrast, both ${\mathcal C}_{\textrm {N\'eel}}^x$ and
${\mathcal C}_{\textrm {N\'eel}}^y$ exist for TXYM with XZX-YZY type of interaction,
for arbitrary $\gamma$, and as a consequence,
bicritical points  exist for any values of $\gamma$.
However, the values
of ${\mathcal C}_{\textrm{N\'eel}}^x(n)$ and
${\mathcal C}_{\textrm{N\'eel}}^y(n)$ for all these
models vanish for arbitrary $n$ when $T>0$, 
which is in accordance with the Mermin-Wagner theorem
which states that continuous symmetries of a
system with short-range interactions
cannot be broken spontaneously at nonzero temperatures 
in dimensions two or less\cite{MW}. 

\subsection{Topological phases} 
\label{Topology}
Bloch Hamiltonian $\mathcal H({k})$ preserves
particle-hole, and chiral symmetries as it holds the following 
relations,
\[\left\{\begin{array}{l}
\mathcal P \mathcal H({k}) \mathcal P^{-1}=-\mathcal H(-{k}),\\ [0.4em]
\mathcal C  \mathcal H({k}) \mathcal C^{-1}=-\mathcal H({k}),
\end{array}\right. \]
respectively. Here $\mathcal P=\sigma_x\mathcal K$, 
$\mathcal C=\sigma_x$ and $\mathcal K$ are the particle-hole, 
chiral and complex conjugation operators, respectively. 
Conservation of the two symmetries corresponds to
that of the TRS and they altogether constitute the BDI class.
Different topological phases have been characterized by 
definite values of the winding number, which is defined by 
\[\nu=\frac{1}{2\pi}\oint_C \left(\boldsymbol {\hat g}(k)\times
  \frac{d}{dk}\,\boldsymbol {\hat g}(k)\right)\!d{k},\]
where $\boldsymbol {\hat g}(k)
=\boldsymbol g'(k)/|\boldsymbol g'(k)|$,
and $\boldsymbol g'=(g'_x,g'_y,0)$, with $g'_x=g_z$, and $g'_y=g_y$. 
The off-diagonal Hamiltonian matrix,
$\mathcal H'(k)=\boldsymbol g'\cdot \boldsymbol \sigma$,
is related to $\mathcal H(k)$
under the unitary transformation,
$\mathcal H'(k)=\mathcal U \mathcal H(k)\mathcal U^\dag$,
with $\mathcal U=1/\sqrt 2(I_2-i\sigma_y)$, and $I_2$ is
the $2\times 2$ identity matrix. 
$C$ is a closed curve in the $g'_x$-$g'_y$ plane. 
Winding number counts 
the number of winding around the origin of the
$g'_x$-$g'_y$ plane, while 
it will be considered as positive when the
curve $C$ goes across the counter clockwise direction.
This formulation is valid if $\mathcal H(k)$ preserves the
chiral symmetry. In this case, $\nu$ will be equal to
the Pancharatnam-Berry phase in unit of the angle
$\pi$\cite{Pancharatnam,Berry}.

Topological phase diagrams have been shown in the third
vertical panel of 
Fig. \ref{magnetic-topology} for five different values: 
\(\gamma=-2\, ({\rm c}),\,-1\,({\rm f}),\,0\,({\rm i}),\,
1\,({\rm l}),\,2\,({\rm o})\). Each horizontal panel
demonstrates that there is a faithful coexistence between
the magnetic and topological phases. It means magnetic
phase with $x\,(y)$-component of
staggered correlation,  ${\mathcal C}_{\textrm {N\'eel}}(n)$ overlaps
with topological superconducting phase with $\nu=1\,(-1)$. The remaining 
trivial topological phase ($\nu=0$) coexist with the
phase of no magnetic LRO. Location of bicritical points are
shown by the circles. 
The nontrivial phase is associated with the symmetry protected
zero-energy states. 

 In order to understand the density plots in Fig 
 \ref{magnetic-topology}, magnitude of the correlation functions,
${\mathcal C}_{\textrm{N\'eel}}^x$ and
 ${\mathcal C}_{\textrm{N\'eel}}^y$ for a particular case
when $\gamma=1$
as shown in Fig \ref{magnetic-topology}
(j) and (k), are described. 
They show that both ${\mathcal C}_{\textrm{N\'eel}}^x$ and
${\mathcal C}_{\textrm{N\'eel}}^y$ vanish when $h/J>+1$ and
$h/J<-1$. It means the system undergoes phase transition to the
disordered phase over both the lines $h/J=+1$ and
$h/J=-1$. However, in the intermediate regime,
$-1<h/J<1$, the phases are characterized by either 
${\mathcal C}_{\textrm{N\'eel}}^x \ne 0$ or
${\mathcal C}_{\textrm{N\'eel}}^y\ne 0$ in separate regions. 
More precisely, ${\mathcal C}_{\textrm{N\'eel}}^x \ne 0$, and
${\mathcal C}_{\textrm{N\'eel}}^y =0$
when $hJ'>-1$, as shown in Fig
\ref{magnetic-topology} (j). 
On the other hand, 
${\mathcal C}_{\textrm{N\'eel}}^y \ne 0$
but ${\mathcal C}_{\textrm{N\'eel}}^x = 0$, when $h J'<-1$, 
 as shown in Fig \ref{magnetic-topology} (k).
It means phase transition occurs over the lines
$h J'=- 1$.

Identically, the topological phase diagram shown
in Fig \ref{magnetic-topology} (l)
demonstrates that nontrivial phase with $\nu=+1$ exists  
when $h J'>-1$, and that with
$\nu=-1$ exists  when $h J'<-1$. 
 As a consequence, faithful coexistence
between the magnetic phase ${\mathcal C}_{\textrm{N\'eel}}^x \ne 0$, and
${\mathcal C}_{\textrm{N\'eel}}^y =0$ with topological
phase with $\nu=+1$, as well as that between 
the magnetic phase ${\mathcal C}_{\textrm{N\'eel}}^x = 0$, and
${\mathcal C}_{\textrm{N\'eel}}^y  \ne 0$ with topological
phase with $\nu=-1$ exists.

Similar arguments hold for other values $\gamma=-2,-1,0,2$, 
as shown in Fig \ref{magnetic-topology}  [(a), (b), (c)], [(d), (e), (f)],
[(g), (h), (i)], and  [(m), (n), (o)].
These results show that for arbitrary $\gamma$,
the magnetic phase with ${\mathcal C}_{\textrm{N\'eel}}^x \ne 0$, and
${\mathcal C}_{\textrm{N\'eel}}^y =0$, coexists with topological
phase with $\nu=+1$ when   $h J'>-\gamma$.
Identically, the magnetic phase with ${\mathcal C}_{\textrm{N\'eel}}^x = 0$, and
${\mathcal C}_{\textrm{N\'eel}}^y  \ne 0$, coexists with topological
phase with $\nu=-1$, when $h J'<-\gamma$. 

Existence of topologically trivial and nontrivial phases had been
experimentally demonstrated 
in terms of the absence and presence of zero-energy states in a
number of ways with classical and quantum systems.
Classically it is demonstrated in acoustic\cite{Wen},
mechanical\cite{Merlo}, photonic\cite{Amo}, plasmonic\cite{Notomi}
systems as well as in microwave resonator\cite{Poli},
and thermal lattices\cite{Hu}. In these
systems edge states of SSH models have been detected. 
 In quantum approach with Rydberg atoms, an artificial system of 14 sites 
 with $^{87}$Rb atoms bound together by interatomic dipolar interaction
 has been setup. The system realizes a bosonic version of
 the SSH model, in which the hard-core bosonic constraint
 is maintained by the infinite onsite interaction energy
 in order to impose the fermionic Pauli exclusion principle
 in the lattice points\cite{Browaeys}.
The similar real-space lattice potential mimicking the
 SSH Hamiltonian has been recently realized with $^{84}$Sr atoms\cite{Kanungo}.
 The existence of zero energy edge states has been successfully demonstrated
 in both the experiments.
Existence of edge states and dynamics of SSH model 
using the $^{87}$Rb atoms in a momentum-space lattice
has been observed\cite{Gadway}.

It is expected that the topological phase in TXYM with XZX$-$YZY term
can be experimentally verified with the detection of zero energy edge states.
Effect of magnetic field on the spin models have been
studied in different contexts\cite{LSM,Katsura,Barouch1,Barouch2,Barouch3,Pfeuty,Mikeska,Bose,Ghosh1,Ghosh2,Ghosh3,Paul}.
Under the JW transformation, this spin chain has been converted
to spinless superconducting model where magnetic field
acts like the onsite potential which plays crucial role in
QPTs\cite{Kitaev}. In the topological superconducting state
zero energy BdG quasiparticles are interpreted as the Majorana
bound states.
Two different types of Majorana edge modes
are realizable in two extreme limits of the
nontrivial topological phase and these are identical
for the Hamiltonians, TXYM with XZX$\pm$YZY terms\cite{Rakesh2}. 
Several protocols have been proposed
in terms of dynamic current susceptibility,
strong-field dynamics, Andreev bound states, etc, in order to
detect this typical feature \cite{Trif,Liu,Grass}.
In this model, magnetic field is found to oppose the
LRO or the existence of topological phase, in the same way
like the TIM and TXYM.
However, there are several systems in which magnetic field
is found essential for exhibiting the topological phases
\cite{Joshi,Moumita1,Sil,Moumita2,Owerre2,Moumita3,Owerre3,Moumita4}.
\section{Concurrence, mutual information and quantum discord}
\label{entanglement}
\begin{figure}[h]
\psfrag{a}{(a)}
\psfrag{b}{(b)}
\psfrag{C2}{$dC(2)/dh$}
\psfrag{T}{$J\!=\!1,J'\!=\!3,\gamma\!=\!1$}
\psfrag{h}{$h/J$}
\psfrag{d}{Disordered}
\psfrag{Cx}{${\mathcal C}_{\textrm{N\'eel}}^x\ne 0$}
\psfrag{Cy}{${\mathcal C}_{\textrm{N\'eel}}^y\ne 0$}
\psfrag{Jp}{$J'/J$}
\psfrag{0}{0}
\psfrag{1.0}{$1.0$}
\psfrag{-1.0}{$-1.0$}
\psfrag{2}{2}
\psfrag{-1.0}{$-1.0$}
\psfrag{-3}{$-3$}
\psfrag{-2}{$-2$}
\psfrag{-4}{$-4$}
\psfrag{3}{3}
\psfrag{4}{4}
\psfrag{-0.5}{$-0.5$}
\psfrag{-2.0}{$-2.0$}
\psfrag{-1.5}{$-1.5$}
\psfrag{1.5}{$1.5$}
\psfrag{2.0}{$2.0$}
\psfrag{0.0}{$0.0$}
\psfrag{0.5}{$0.5$}
\psfrag{-0.8}{$-0.9$}
\psfrag{-0.9}{$-0.9$}
\psfrag{0.15}{$0.15$}
\psfrag{0.20}{$0.20$}
\psfrag{0.25}{$0.25$}
\includegraphics[width=250pt]{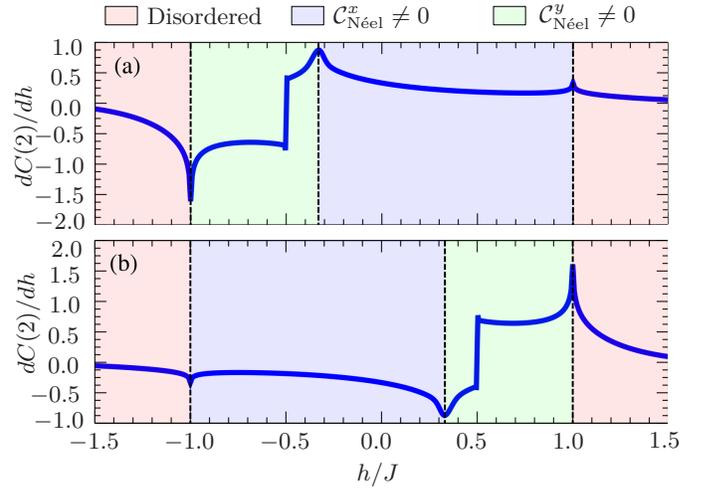}
\caption{Variation of first order derivative of concurrence, $C(2)$ 
  with respect to $h/J$, for $\gamma=1$ (a), $\gamma=-1$ (b),
  when $J'=3$. Different phases are shown in different colors. Vertical
dashed lines indicate the phase transition points.}
 \label{concurrence}
\end{figure}

QPTs can be identified by studying the
properties of entanglement measures in the vicinity of
transition points. The correlation of quantum state
can be quantified in terms of bipartite entanglements.
For two-qubit system one such measure is the concurrence,
which is applicable for both
pure and mixed states\cite{Wooters,Moumita5}. 
In a spin-1/2 system,
two-site reduced density matrix for spins of two arbitrary
positions, $i$  and $j$ is given by\cite{Lin}
\be
\rho_{ij}=\frac{1}{4}\!
+\!\sum_\alpha \!\left(\langle S_i^\alpha \rangle  S_i^\alpha
\!+\!\langle S_j^\alpha \rangle  S_j^\alpha\right)\!+\!
\sum_{\alpha \beta}\langle S_i^\alpha S_j^\beta \rangle S_i^\alpha S_j^\beta.
\ee
The reduced density matrix in terms of the standard
two-site two-qubit basis states 
$\{|11\rangle,\,|10\rangle,\,|01\rangle,\,|00\rangle\}$
assumes a simpler form\cite{Ali}.
If the two sites are separated by 
distance of $n$ units,
\be \rho(n)=\left(
\begin{array}{cccc}u_+(n)&0&0&v_-(n)\\[0.4em]
  0&w(n)&v_+(n)&0\\[0.2em]
  0&v_+(n)&w(n)&0\\[0.2em]
   v_-(n)&0&0&u_-(n)\end{array}
\right),\label{dm}\ee
where the non-zero matrix elements are given
in terms of magnetization and correlations as
\be\left\{
\begin{aligned}
&u_\pm(n)=1/4\pm m_z +{\mathcal C}_{\textrm{N\'eel}}^z(n),\\[0.4em]
&w(n)=1/4-{\mathcal C}_{\textrm{N\'eel}}^z(n),\\[0.4em]
&v_\pm(n)={\mathcal C}_{\textrm{N\'eel}}^x (n)\pm {\mathcal C}_{\textrm{N\'eel}}^y(n).
\end{aligned}\right.
\ee
Finally, concurrence between two spins separated by
$n$, $C(n)$ is given by $C(n)={\rm Max}\{C_1(n),C_2(n)\}$, where
\be\left\{
\begin{aligned}
&C_1(n)=2(|v_+(n)|-\sqrt{u_+(n)u_-(n)}),\\[0.4em]
&C_2(n)=2(|v_-(n)|-w(n)).
\end{aligned}\right.
\ee
Variation of the derivative of concurrence between two spins separated by
$n=2$, $dC(2)/dh$ is shown in Fig \ref{concurrence}
for $\gamma=1$ (a), $\gamma=-1$ (b), when $J'=3$. 
The derivative $dC(2)/dh$ exhibits distinct peaks at the transition points.
It reveals that the concurrence exhibits the
symmetry, $C(n=2,J,\pm \gamma,J',\pm h)
=C(n=2,J,\mp \gamma, J',\mp h)$, which 
corresponds to the relation Eq (\ref{symmetry}f).
 Sharp jumps are noted at $h/J=-0.5$, and $h/J=0.5$,
when $\gamma=1$, and $\gamma=-1$, respectively.
It attributes to the fact that $C(2)$ changes sign there. 

Other entanglement measures like
mutual information and quantum discord
have been employed here in order to
study the character of quantum correlations\cite{Amico}. 
Correlation can be divided into classical and quantum origins,
where the total correlation can be measured by 
quantum mutual information (MI). In a bipartite
system MI ($Mi$) between two subsystems $A$ and
$B$ can be defined as 
\be
Mi(\rho_{AB})=S(\rho_{A})+S(\rho_{B})-S(\rho_{AB}),
\label{MI}
\ee
where $\rho_{AB}$ is the density matrix of
the whole system, while 
$\rho_{A}$ and $\rho_{B}$ are the reduced density
matrices of the subsystems $A$ and $B$, where 
$S(\rho)=-Tr\{\rho\log_{2}\rho\}$, 
is the von Neumann entropy for the density matrix, $\rho$. 

\begin{figure}[h]
\psfrag{a}{\color{white}(a)}
\psfrag{b}{\color{white}(b)}
\psfrag{D}{\color{white} Disordered}
\psfrag{A}{\color{white} ${\mathcal C}_{\textrm{N\'eel}}^x\ne 0$}
\psfrag{B}{\color{white}${\mathcal C}_{\textrm{N\'eel}}^y\ne 0$}
\psfrag{h0}{$h_{\rm z}\!=\!0$}
\psfrag{T}{$J\!=\!1,J'\!=\!3,\gamma\!=\!1$}
\psfrag{h}{$h/J$}
\psfrag{jp}{$J'/J$}
\psfrag{0}{0}
\psfrag{1.0}{$1.0$}
\psfrag{-1.0}{$-1.0$}
\psfrag{2}{2}
\psfrag{-1.0}{$-1.0$}
\psfrag{-3}{$-3$}
\psfrag{-1}{$-1$}
\psfrag{-2}{$-2$}
\psfrag{-4}{$-4$}
\psfrag{1}{1}
\psfrag{2}{2}
\psfrag{3}{3}
\psfrag{4}{4}
\psfrag{-0.5}{$-0.5$}
\psfrag{-1.5}{$-1.5$}
\psfrag{-2.0}{$-2.0$}
\psfrag{1.5}{\hskip 0.0 cm $1.5$}
\psfrag{0.0}{$0.0$}
\psfrag{0.5}{$0.5$}
\psfrag{0.8}{$0.8$}
\psfrag{-0.9}{$-0.9$}
\psfrag{1.6}{$1.6$}
\psfrag{1.4}{$1.4$}
\psfrag{1.2}{$1.2$}
\psfrag{0.2}{$0.2$}
\psfrag{0.1}{$0.1$}
\psfrag{0.3}{$0.3$}
\psfrag{0.4}{$0.4$}
\psfrag{0.00}{$0.00$}
\psfrag{0.01}{$0.01$}
\psfrag{0.05}{$0.05$}
\psfrag{0.02}{$0.02$}
\psfrag{0.03}{$0.03$}
\psfrag{0.04}{$0.04$}
\hskip -0.7 cm
\includegraphics[width=260pt,angle=0]{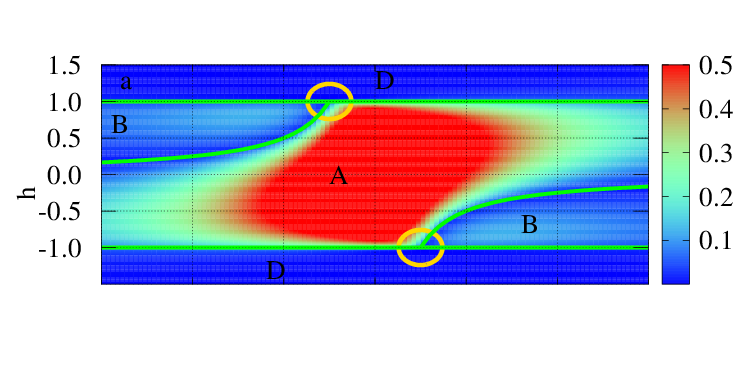}
\vskip -1.8 cm
\hskip -0.7 cm
\includegraphics[width=260pt,angle=0]{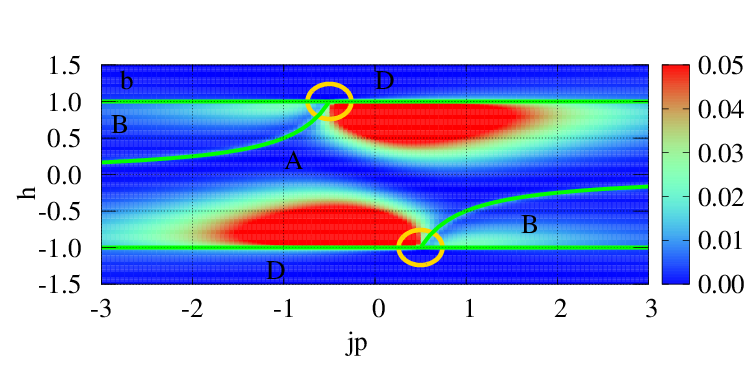}
\caption{Density plot for mutual information (a) and quantum discord (b) 
  in the $h$-$J'$ plane is shown for $\gamma=1/2$, when $n=1$. 
  Upper and lower bicritical
  points are enclosed within yellow circle. }
 \label{mutual-information}
\end{figure}

\begin{figure}[h]
\psfrag{a}{\color{yellow}(a)}
\psfrag{b}{\color{yellow}(b)}
\psfrag{D}{\color{white} Disordered}
\psfrag{A}{\color{white} ${\mathcal C}_{\textrm{N\'eel}}^x\ne 0$}
\psfrag{B}{\color{white}${\mathcal C}_{\textrm{N\'eel}}^y\ne 0$}
\psfrag{M}{\color{blue}$m_z$}
\psfrag{h}{\hskip 0.0 cm $h$}
\psfrag{h0}{$h_{\rm z}\!=\!0$}
\psfrag{p1}{$J\!=\!1,J'\!=\!1,\gamma\!=\!2.0$}
\psfrag{p2}{$J\!=\!1,J'\!=\!1,\gamma\!=\!-2.0$}
\psfrag{h}{$h/J$}
\psfrag{jp}{$J'/J$}
\psfrag{0}{0}
\psfrag{1.0}{$1.0$}
\psfrag{-1.0}{$-1.0$}
\psfrag{2}{2}
\psfrag{-1.0}{$-1.0$}
\psfrag{-3}{$-3$}
\psfrag{-1}{$-1$}
\psfrag{-2}{$-2$}
\psfrag{-4}{$-4$}
\psfrag{1}{1}
\psfrag{2}{2}
\psfrag{3}{3}
\psfrag{4}{4}
\psfrag{-0.5}{$-0.5$}
\psfrag{-1.5}{$-1.5$}
\psfrag{1.5}{\hskip 0.0 cm $1.5$}
\psfrag{0.0}{$0.0$}
\psfrag{0.5}{$0.5$}
\psfrag{0.8}{$0.8$}
\psfrag{-0.9}{$-0.9$}
\psfrag{2.0}{$2.0$}
\psfrag{2.5}{$2.5$}
\psfrag{3.0}{$3.0$}
\psfrag{1.6}{$1.6$}
\psfrag{1.4}{$1.4$}
\psfrag{1.2}{$1.2$}
\psfrag{0.2}{$0.2$}
\psfrag{0.4}{$0.4$}
\psfrag{0.6}{$0.6$}
\psfrag{0.00}{$0.00$}
\psfrag{0.01}{$0.01$}
\psfrag{0.05}{$0.05$}
\psfrag{0.02}{$0.02$}
\psfrag{0.03}{$0.03$}
\psfrag{0.04}{$0.04$}
\hskip -0.7 cm
\includegraphics[width=260pt,angle=0]{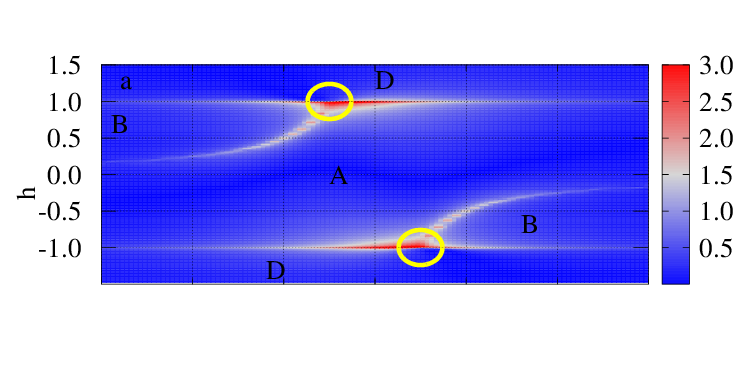}
\vskip -1.8 cm
\hskip -0.7 cm
\includegraphics[width=260pt,angle=0]{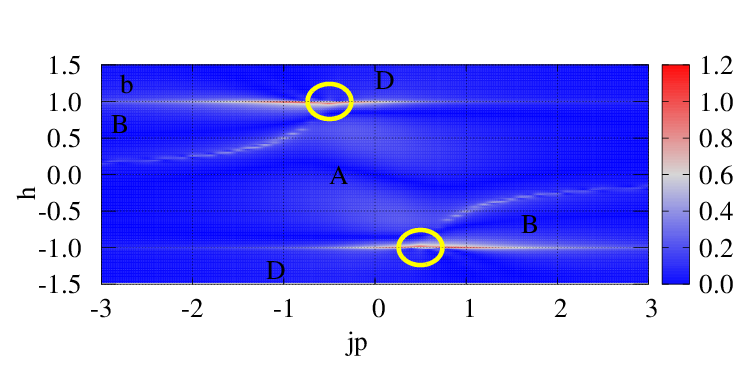}
\caption{Density plot for absolute value of first derivative of
  mutual information (a) and quantum discord (b) 
  in the $h$-$J'$ plane is shown for $\gamma=1/2$, when $n=1$. 
  Upper and lower bicritical
  points are enclosed within yellow circle. }
 \label{derivative-mutual-information}
\end{figure}

In this two-qubit system, MI can be expressed as
\be
Mi(\rho_{AB})=\sum_{m=1}^4d_m \log_{2}{d_m}-\sum_{\alpha=\pm}d_\alpha \log_{2}{(d_\alpha/2)},
\ee
where, $d_m,\,m=1,2,3,4$, are the eigenvalues of
$\rho_{AB}$, while  $d_\pm=1\pm(u_+(n)-u_-(n))$.
MI quantifies the total correlation, whereas quantum
correlation for a bipartite system can be measured  
in terms of quantum discord (QD)\cite{Ollivier,Luo,Sarandy}.
In this bipartite system, QD ($Qd$) can be expressed as
\be
Qd(\rho_{AB})={\rm Min}\,\{D_1,\,D_2\},
\ee
where
\bea
D_1\!&=&D-\sum_{\alpha=\pm}\left[
  u_\alpha\log_{2}{\left(\frac{u_\alpha}{u_\alpha+w}\right)}
-w\log_{2}{\left(\frac{w}{u_\alpha+w}\right)}\right],\nonumber\\[0.4em]
D_2\!&=&D-\sum_{\alpha=\pm}\Delta_\alpha\log_{2}{\Delta_\alpha},\nonumber
\eea
and
\be
\begin{aligned}
&D=\sum_{m=1}^4d_m \log_{2}{d_m}-\sum_{\alpha=\pm}\frac{d_\alpha}{2} \log_{2}{\left(\frac{d_\alpha}{2}\right)},\\[0.4em]
&\Delta_\pm=\frac{1}{2}(1\pm \Delta),\\[0.4em]
&\Delta=\sqrt{(u_+(n)-u_-(n))^2+4(|v_+(n)|+|v_-(n)|)^2}.
\end{aligned} \nonumber
\ee

Variation of MI (a) and QD (b) in the $h$-$J'$ plane
are shown in Fig \ref{mutual-information},
where density plots is drawn for $\gamma=1/2$. 
Upper and lower bicritical points are enclosed within yellow circles.
Different magnetic phases are noted there. 
They reveals that values of both MI and QD vanishes
over the phase transition lines. Both are zero beyond the
Ising transition lines ($h/J>1$, and $h/J<-1$). 
In addition, MI vanishes in the region around the point
$h=0,\,J'=0$, where QD is maximum.
Similarly, variation of absolute values of the
derivatives of MI and QD in the $h$-$J'$ plane
are shown in the density plots, Fig \ref{derivative-mutual-information} (a) and
(b), respectively for $\gamma=1/2$. 
Peaks of the absolute values of the derivatives
$|dMi/dh|$ and $|dQd/dh|$
mark the location of QPTs which perfectly overlap on the
Ising and anisotropy transition lines
along with the pair of bicritical points shown within the
circles. Different magnetic phases have been identified. 
\section{Discussion}
\label{Discussion}
Many years before Suzuki proposed a class of generalized XY model
in the presence of terms with more than two spins
interactions those are solvable
under JW transformations\cite{Suzuki}.
These interactions are chosen in such a way that they
essentially lead to single fermionic terms in the
transformed Hamiltonians and it ultimately triggers a series of
investigations with a variety of three-spin terms.
In this work, AFM TXYM Hamiltonian in the presence of
XZX$-$YZY type of interactions has been considered
and extensive phase diagrams for magnetic and topological
properties have been prepared.
Magnetic LROs are found and faithful coexistence between
magnetic and topological orders are observed. 
Quantum correlations
in terms of several entanglement measures have been studied. 
Real materials exhibiting the presence of 
three-spin interaction has been discovered before.
In the magnetic compound CsMn$_{x}$Mg$_{1-x}$Br$_3$, the Mn$^{2+}$ ions
constitute one-dimensional isotropic
Heisenberg antiferromagnet when $x=1$. 
However, upon doping with Mg replacing the Mn atoms, and when $x=0.28$,
existence of biquadratic three-spin interaction has been predicted.
The inelastic-neutron scattering experiment has revealed the presence of
magnetic LRO in CsMn$_{0.28}$Mg$_{0.72}$Br$_3$\cite{Falk}. 
In order to engineer three-spin interaction of any specific forms 
several attempts have been made. The effective Hamiltonian comprising
three-spin term has been demonstrated using tailored ion-laser interactions
in a linear Paul trap enhanced by Molmer-Sorensen scheme. In this
method QPT in one-dimensional spin systems has
been demonstrated\cite{Martin-Delgado}.
It is expected that magnetic phases for this model may be
verified in this route. 
 \section{ACKNOWLEDGMENTS}
  RKM acknowledges the DST/INSPIRE Fellowship/2019/IF190085.
  \section{Data availability statement}
  All data that support the findings of this study are included within the article.
  \section{Conflict of interest}
  Authors declare that they have no conflict of interest.  
   \appendix


\end{document}